# Terahertz electric-field driven dynamical multiferroicity in SrTiO$_3$

M. Basini[1], M. Pancaldi[2,3], B. Wehinger[2†], M. Udina[4], T. Tadano[5], M. C. Hoffmann[6], A. V. Balatsky[2,7,8], S. Bonetti[1,2,*]

1. Department of Physics, Stockholm University, 106 91 Stockholm, Sweden
2. Department of Molecular Sciences and Nanosystems, Ca' Foscari University of Venice, 30172 Venice, Italy
3. Elettra-Sincrotrone Trieste S.C.p.A., 34149 Basovizza, Italy
4. Department of Physics and ISC-CNR, "Sapienza" University of Rome, 00185, Rome, Italy
5. Research Center for Magnetic and Spintronic Materials, National Institute for Materials Science, Tsukuba 305-0047, Japan
6. Linac Coherent Light Source, SLAC National Accelerator Laboratory, Menlo Park, CA 94025, USA
7. NORDITA, 106 91 Stockholm, Sweden
8. Department of Physics, University of Connecticut, CT 06268, United States

†Present address: European Synchrotron Radiation Facility, 38043 Grenoble, France

* corresponding author: stefano.bonetti@fysik.su.se

**The emergence of collective order in matter is among the most fundamental and intriguing phenomena in physics. In recent years, the ultrafast dynamical control and creation of novel ordered states of matter not accessible in thermodynamic equilibrium is receiving much attention[1-6]. Among those, the theoretical concept of dynamical multiferroicity has been introduced to describe the emergence of magnetization by means of a time-dependent electric polarization in non-ferromagnetic materials[7,8]. In simple terms, a large amplitude coherent rotating motion of the ions in a crystal induces a magnetic moment along the axis of rotation. However, the experimental verification of this effect is still lacking. Here, we provide evidence of room temperature magnetization in the archetypal paraelectric perovskite SrTiO$_3$ due to this mechanism. To achieve it, we resonantly drive the infrared-active soft phonon mode with intense circularly polarized terahertz electric field, and detect a large magneto-optical Kerr effect. A simple model, which includes two coupled nonlinear oscillators whose forces and couplings are derived with *ab-initio* calculations using self-consistent phonon theory at a finite temperature[9], reproduces qualitatively our experimental observations on the temporal and frequency domains. A quantitatively correct magnitude of the effect is obtained when one also considers the phonon analogue of the reciprocal of the Einsten – de Haas effect, also called the Barnett effect, where the total angular momentum from the phonon order is transferred to the electronic one. Our findings show a new path for designing ultrafast magnetic switches by means of coherent control of lattice vibrations with light.**

It is now established that noncollinear magnetic order in magnetic insulators can induce an electric polarization along the axis of the spiral spin structure[10]. The fundamental microscopic mechanism involved is the Dzyaloshinskii-Moriya interaction (DMI), which has the form $S_i \times S_j$ for spins $S$ at sites $i$ and $j$, and which is known to promote noncollinear magnetic order[11-13]. By including full relativistic corrections owing to the spin-orbit coupling, the DMI leads to a polarization $P$ of the form $P \sim e_{ij} \times (S_i \times S_j)$, where $e_{ij}$ is the unit vector between sites $i$ and $j$, as shown in Ref. [10]. On the other hand, from symmetry considerations, the permutation of space and time, and of electric and magnetic fields, a magnetization $M$ of the form $M \sim P \times \partial_t P$ is expected to appear in the presence of a time-dependent polarisation[7]. Within a classical picture, the motion of ions in a closed-loop induces an orbital magnetic moment, which is in the order of the nuclear magneton $\mu_N \approx 10^{-3} \mu_B$ per unit cell (u.c.). Inducing such ionic motion is the main idea of the experiment presented here and illustrated in Fig. 1. Starting from a static configuration which produces no net magnetic moment in the unit cell of strontium titanate SrTiO$_3$ (STO), the application of a circularly polarized terahertz electric field induces a coherent rotation of the ions generating $M$. The sign of this magnetization is controlled by reversing the helicity of the terahertz field.

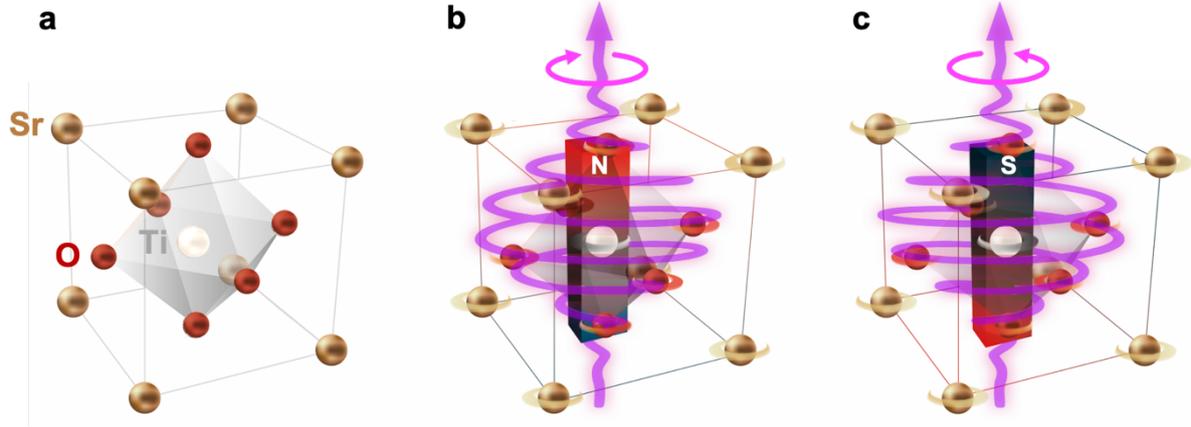

**Figure 1 | Schematic of the experimental realization of dynamical multiferroicity**. (a) SrTiO$_3$ unit cell in the absence of a terahertz electric field. When a circularly polarized terahertz field pulse drives a circular atomic motion, dynamical multiferroicity is expected to create a net magnetic moment in the unit cell, with (b) a north pole up for a pulse which is left-handed and (c) a south pole up for a pulse which is right-handed.

STO is a paraelectric diamagnetic material, with a cubic perovskite structure at room temperature. It provides an interesting case for probing electric field driven dynamics, having several zone-center phonon modes in a frequency range accessible to modern terahertz sources[14-16]. The polar ferroelectric phonon mode can be driven circularly due to its three-fold degeneracy in the cubic crystal structure at room temperature, and it is infrared active. This mode softens significantly in frequency from $\nu \approx 2.7$ THz at $T = 300$ K to $\nu \approx 0.2$ THz at $T = 5$ K[17,18], but it does not establish long-range ferroelectric order: quantum fluctuations prevent the system from remaining in one of its electronic potential minima even at zero temperature. The mode-selective drive by a strong THz field allows for excitation of the ferroelectric soft mode into highly anharmonic regimes, enabling coupling to modes with different symmetry and of higher frequency, which have recently been probed by means of ultrafast x-ray diffraction[19].

Even though such large amplitude motion of the ferroelectric mode driven by resonant linearly polarized terahertz fields was unambiguously demonstrated, its response to circularly polarized narrow-band THz pulses and the potential onset of a magnetization via dynamical multiferroicity are still unexplored. Here, we attempt this investigation by using terahertz pulses centred at 3 THz with a bandwidth of 0.5 THz, generated with a table-top source, as described in the Methods. The circular polarization is obtained with a quarter-wave plate designed for the 3 THz (100 μm) radiation. This circularly polarized pump beam, propagating along the crystallographic [001] direction of a 500 μm thick STO crystal, can be seen as a superposition of two perpendicular pulses linearly polarized along the [100] and [010] directions, phase-shifted by $\pi/2$. The sample is mounted tilted at 45 degrees w.r.t. the free space propagation direction to measure in reflection geometry. However, the large dielectric constant ($|\tilde{\varepsilon}| \sim 100$) of STO at the resonant frequency of the soft phonon mode causes such a large refraction of the beam that the propagation within the crystal is always orthogonal w.r.t. the sample surface. The ferroelectric phonon mode is thus driven along two of the three degenerate eigenvectors with the same phase shift, resulting in circular ionic motion. The polar nature of the phonon implies a polarization ***P*** orthogonal to its time derivative $\partial_t \boldsymbol{P}$ during the whole dynamics, thus maximising the cross product expected to give rise to the magnetization. In order to probe possible magnetic signals driven by the terahertz-induced coherent phonon dynamics, we used the time-resolved magneto-optical Kerr effect (MOKE), which measures the rotation of the polarization of a probe pulse reflected by a magnetized material[20]. Both the pump and probe pulses are generated from the same amplified laser system, so that the comparatively slower terahertz electric field is phase-stable within the probe resolution. The MOKE has been shown to be sensitive to an effective magnetic field driven by optical phonons in antiferromagnetic ErFeO$_3$ at low temperatures[21]. In that case, the amplification of an already existing coherent spin precession resonant with the induced magnetic field by the phonon motion was used as an indirect proof of the existence of such field. In the following, we show that magnetic order can be induced via dynamical multiferroicity also in non-magnetic systems where there are no underlying magnetic excitations.

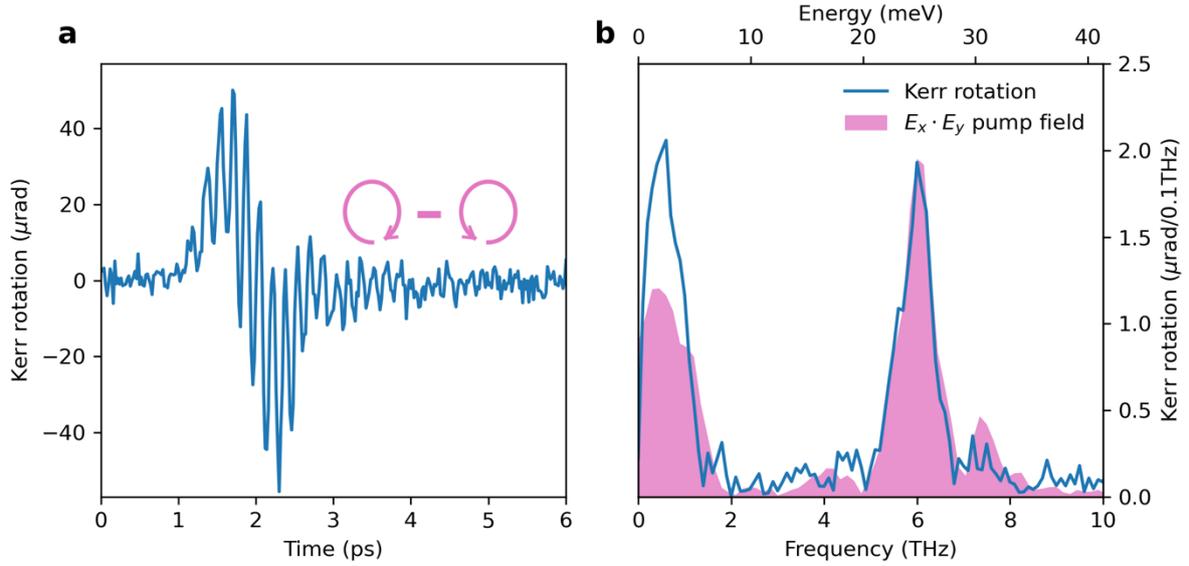

**Figure 2 | Experimental detection of the time-resolved Kerr rotation.** Measured Kerr rotation as a function of pump-probe delay for 400 nm probe pulses after excitation via THz pump fields. The probe polarisation is parallel to the [100] crystal axis and the sample tilted at an angle of 45 degrees (see Methods). (a) Difference of the responses from circularly polarised fields with opposite helicities. (b) FFT of the time trace in (a) together with the spectrum of the product of the two components $E_x$ and $E_y$ of the incident circular pump field (see Methods).

The temporal evolution of the Kerr rotation of a 400 nm, 40 fs long probe pulse as a function of the time delay with respect to terahertz pump pulses of approximately 230 kV/cm in amplitude, is shown in Fig. 2a. We plot the difference signals measured using circularly polarized fields with opposite helicities. The signal exhibits one cycle of a slower oscillation at a frequency $\omega_-$ modulated by a faster oscillating component at a frequency $\omega_+$. In Fig. 2b we plot the calculated fast Fourier transform (FFT) of the time-trace of Fig. 2a. Such analysis clearly reveals a low-frequency component centred at approximately 0.6 THz, and a faster one at around 6 THz. In the same figure, we also show the FFT of the product of the two components $E_x$ and $E_y$ composing the difference of the z-propagating circular pump field (centred at 3 THz, shown in Extended Data Fig. 1), retrieved with an independent electro-optical measurement (see Methods). We normalise such product so that the peaks of the 6 THz components overlap. We will keep an identical approach when evaluating the theoretical response, given that the units of the two measurements are different and scaling factors are arbitrary. We highlight already now that, by aligning the high-frequency peaks, there is a different amplitude in the low frequency peaks, with the measured Kerr rotation being larger than the product of the two field components.

In Fig. 3a we plot the FFT of the measured difference Kerr rotation as a function of temperature, from 160 K to 360 K, a temperature range large enough to move the soft phonon in and off resonance with the driving THz field. We note that the overall response is larger at a temperature of 280 K, and it decreases as the temperature moves away from it, both at lower and higher temperatures. Fig. 3b shows instead the amplitude of the two FFT components of the measured Kerr rotation at $T = 300$ K as a function of the applied terahertz electric field amplitude. For both frequency components, a clear quadratic dependence is observed, with a larger curvature for the lower frequency mode at $\omega_-$ than for the higher frequency one at $\omega_+$.

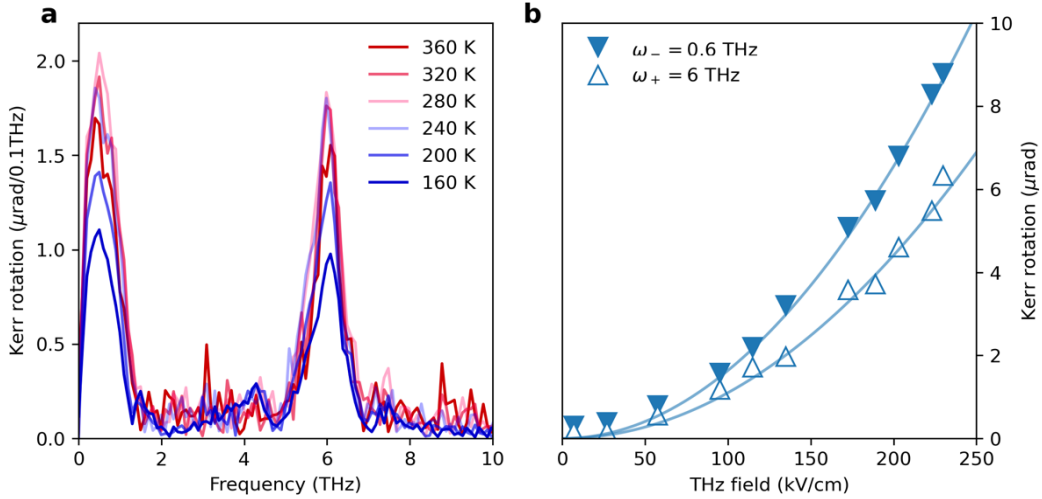

**Figure 3 | Temperature and field dependence of the measured Kerr rotation** (a) FFT of the difference of MOKE response from opposite circularly polarised fields at a few selected temperatures. (b) Symbols: THz electric field dependence of the amplitude of the two main spectral peaks. Solid lines: quadratic fit to the data at $\omega_-$ and $\omega_+$.

We now start to discuss the possible origin of the two peaks in the difference signal. Their frequency is consistent with the prediction of dynamical multiferroicity[7], with a peak $\omega_+$ at twice the soft phonon frequency, and one at $\omega_- \approx 0$. In the original model, no anharmonic effects were included, which could result in a slight frequency shift for the observed dynamics. However, the major issue when comparing the prediction and the experimental evidence, is the relative ratio of the amplitude of the two peaks. From the dynamical multiferroicity theory[7], one would expect the peak at $\omega_+$ to have negligible amplitude, while the experimental data show that they are approximately equal.

Two peaks of comparable amplitudes are on the other hand expected when accounting for the electronic nonlinear response, associated with $\chi^{(3)}$, i.e. the third-order susceptibility tensor[22]. We present a detailed combined theoretical and experimental study on this aspect in a separate work[23], accounting for the symmetry properties of the crystal and for the specific spectral features of the pump pulse. We anticipate here that the theory predicts an effect proportional to the product of the $E_x$ and $E_y$ components of the incident THz field, as we show in Fig. 2b. Briefly, the product of the two THz fields component at a frequency $f$ gives rise to a signal probed by the near infrared field at $2f$ and to a rectified signal, different from zero because of the finite bandwidth of the laser pulse. However, we note that while the shape and amplitude of the peak at $\omega_+$ can be reproduced exactly by the calculations based on the symmetry of the $\chi^{(3)}$ tensor, the experimental peak at $\omega_-$ is higher than what those same calculations predict, and which cannot be accounted for with a different normalization scheme. Hence, the question arises on what fraction of the peak at $\omega_-$ is due to dynamical multiferroicity, and how it can be disentangled from possible $\chi^{(3)}$ contributions. The stronger quadratic dependence of the $\omega_-$ mode on the driving terahertz field, as compared to the one of the $\omega_+$ mode also supports the presence of an underlying extra contribution to the overall sample response.

We notice that the dynamical multiferroicity picture applied to our case would result in a quasi-static magnetic moment whose sign is given by the helicity of the excitation. It is important to stress again that in the original dynamical multiferroicity theory[7], no anharmonic effects were taken into account. On the other hand, the phonon frequency and polarization vectors of $SrTiO_3$ are known to be strongly temperature-dependent due to lattice anharmonicities[24,25]. Therefore, for a quantitative description of the phenomenon, we calculate the effective phonon frequencies and polarization vectors renormalized by the quartic anharmonicity by means of the self-consistent phonon theory[9], which we describe in detail in the Methods section.

We use these calculations to estimate the potential energy landscape for the two orthogonal displacement directions of the soft phonon. Then, we implement the theory of dynamical multiferroicity to compute the expected induced magnetization, which we plot in Fig. 4a and 4b in the time and, respectively, frequency domains. In Fig. 4b, we also show the simulated non-linear optical response of the material, proportional to the product of the $E_x$ and $E_y$. Finally, the expected polarization rotation resulting from the sum of the two effects is shown in the same figures, where we used the identical normalization scheme as for the experimental data, matching the amplitude of the 6 THz mode. We used the computed parameters for all quantities but the soft phonon eigenfrequency and its linewidth, where experimental data is available[17, 18] and more consistent with our own experimental observations. It is well-known that these two quantities are particularly difficult to estimate *ab initio*, returning values within 20%-30% of the experimental observations. The computed magnetic moment is of the order of $10^{-2}$ $\mu_N$ per u.c., much smaller than the one computed in Ref.[8], but consistent with a more realistic atomic displacement, comparable to the one observed experimentally in Ref.[19].

The resemblance between Fig. 4a-b and Fig. 2a-b is remarkable, given that no free adjustable parameters were used in our calculations. In particular, we note that by combining the nonlinear optical response and the dynamical multiferroicity calculations, we can obtain two frequency peaks with relative weights matching the experimental data. This strongly supports the fact that the additional spectral weight at the low frequency peak could be explained with the emergence of a magnetic moment due to dynamical multiferroicity. In order to investigate this further, we plot in Fig. 4c the expected magnetic moment induced via the dynamical multiferroicity mechanism as a function of temperature, again with no free adjustable parameters, and compare it to the measured extra weight of the low frequency peak at different temperatures. The qualitative agreement between calculations and experiment in the temperature-dependent behavior is outstanding, showing a peaked response close to the temperature where the soft phonon is resonant with the driving pump THz field.

The estimated Kerr rotation due to a dynamical multiferroicity effect, calculated using the known Verdet constant for STO, the computed magnetic moment, and the penetration depth of the terahertz radiation (see Methods) is approximately four orders of magnitude larger than the theoretical one, $10^{-1}$ $\mu_B$ vs $10^{-2}$ $\mu_N$ per unit cell. This suggests that, while dynamical multiferroicity can induce an effect qualitatively consistent with our observations, an additional and key mechanism must be in place to explain it quantitatively. We rule out enhancements of the magneto-optical signal similar to those reported in Ref.[26]. While the phenomenology of the inverse Faraday effect is similar, the microscopic mechanism cannot be related, given the three orders of magnitude between their pump photon energy, in the eV range, and the ones used in our work.

Recently, a phonon magnetic moment four orders of magnitude larger than expected from pure phonon motion was observed in a Dirac semimetal[27], where the electron-phonon coupling was deemed to be the underlying fundamental mechanism for the enhancement. In another work[28], the measured *g* factor in PbTe related to a soft phonon at THz frequencies was found to be three orders of magnitude larger than predicted by theory, with the authors inferring that anharmonicity was the reason for the enhancement. A recent theoretical work[29] found that in semimetal and small-gap (tens of meV) insulators, a four orders of magnitude enhancement of the phonon magnetic moment can be expected. None of these results apply directly to our sample, but they all reflect a similarly enhanced magnetic moment when phonons are involved. Finally, *ab initio* calculations have also shown that in STO the coupling strength between the electronic system and the soft phonon at the zone center is very large, of the order of 1 eV[25,30]. We stress, however, that the fundamental physical observable of relevance in this case is not the energy but the angular momentum.

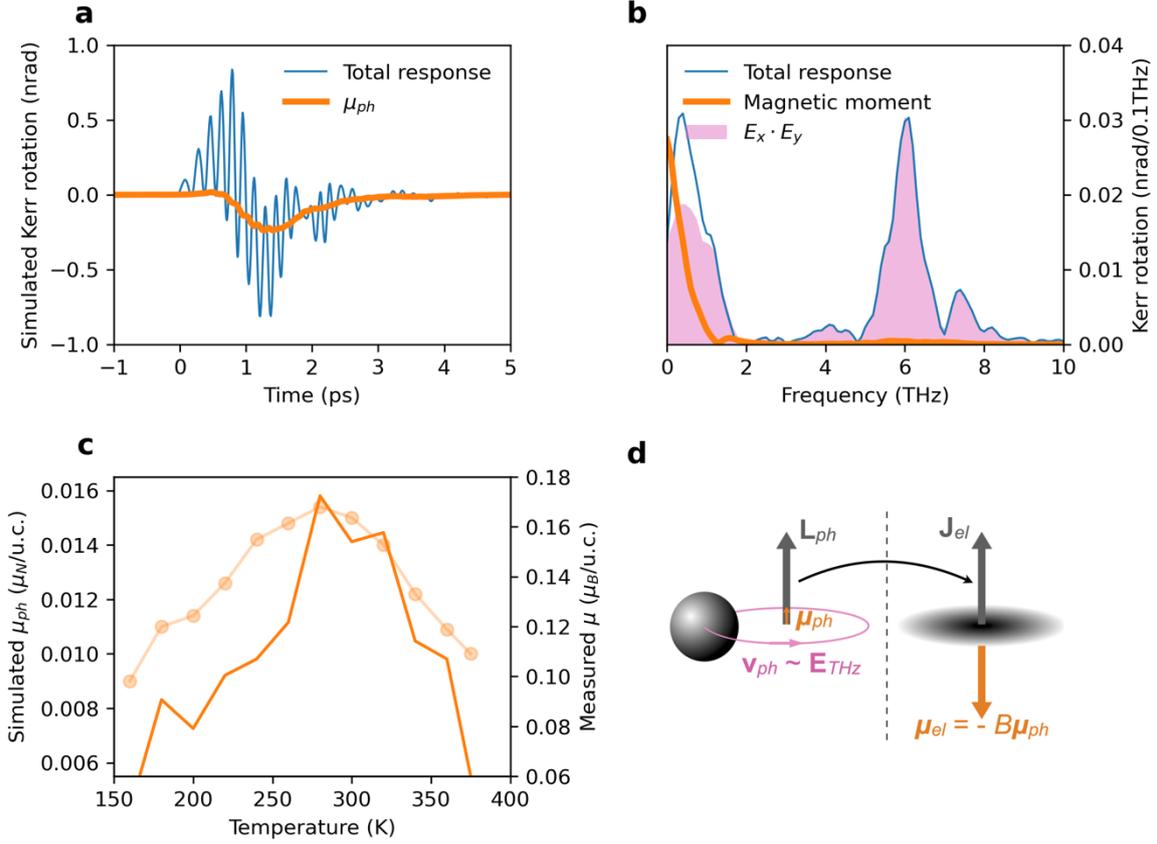

**Figure 4 | Experimental and modelled dynamical multiferroicity.** (a) Time-domain polarization rotation calculated including both the dynamical multiferroicity and the third order electronic response. (b) FFT of the time-domain response, decomposing the spectral features of the two effects. (c) Calculated temperature dependence of the magnetic moment expected from the pure dynamical multiferroicity mechanism (semi-transparent symbols and line), and experimentally extracted spectral weight of the low-frequency peak (solid line). (d) Pictorial representation of the phonon Barnett effect with enhancement factor $B$.

All these observations point to an important gap in our understanding on the transfer of angular momentum from phonon to electrons. It has recently been suggested that in coupled electron-phonon dynamics the reduced mass of the system should be considered[31]. This would intuitively result in a magnetic moment enhanced by a factor of the order of the ratio of the masses of proton and electron, owning to the larger gyromagnetic ratio, as we show pictorially in Fig. 4d. For our case, we observe an enhancement similar to the ratio $2\frac{m_p}{m_e} \approx 4 \cdot 10^3$, where the factor two accounts for the contribution of a neutron for each proton to the nuclear mass. The fundamental mechanism has however never been discussed, and a simple classical analogy of a planetary motion would fail in giving an accurate description. We propose that the observed enhancement of magnetic moment can be thought as the microscopic version of the Barnett effect, i.e., the reciprocal of the Einstein – de Haas effect, where a rigid rotation induces a magnetization in the rotating body. In our case, the body is not rigidly rotating, but it is a coherent phonon motion which induces a transient net angular momentum in the lattice. While a phonon Einstein de Haas effect has been proposed theoretically[32,33,34] and recently observed experimentally[35,36], the experimental realization of the phonon Barnett effect, the transfer of mechanical angular momentum from rotating phonons to the electronic system, has not been reported nor a scheme to detect it yet devised. Due to Onsager reciprocity, such effect must however exist, and we argue that it can be used to explain our results. Our observed enhancement is of the correct order of magnitude, even with the oversimplified assumption that the angular momentum is shared by all electrons equally. We speculate that the angular momentum is likely transferred to a subset of them, possibly the valence electrons, hence the enhancement could be a few times larger. Detailed considerations will be the scope of a future work. We also stress that with the current data we can only assume that the transfer of angular momentum is between the orbital phonon angular momentum $\mathbf{L}_{ph}$ and the total electronic angular momentum $\mathbf{J}_{el}$. The presented results cannot resolve between orbital and spin contributions; the latter could possibly arise if the strong circularly polarized THz field modifies the band structure to such an extent that a spin-polarization is created mediated by spin-orbit coupling, but large signals could be expected due to orbital contributions[37]. We anticipate that our hypothesis can be tested by time-resolved X-ray magnetic circular dichroism experiments at free electron facilities. The total magnetic moment that we measured is of the order of a tenth of a $\mu_B$, which integrated over the probing volume is roughly equivalent to the one of thin ferromagnetic layers that have been already measured at those facilities.

We finally comment on the possible implications of our results. Due to the short phonon lifetime at room temperature, STO does not maintain any memory of the possible magnetic order for longer than a few picoseconds. However, since STO is used commonly as a substrate for the growth of oxide heterostructures, one could conceive of new ways to drive magnetic states at interfaces. At low temperature, intense coherent THz excitation of the soft mode in STO can lead to highly nonlinear phonon responses that overcome the quantum fluctuations creating a ferroelectric order absent at equilibrium[38]. As a next step towards long-living excitation, our study could be extended to the low temperature regime, where a ferroelectric order in doped-STO would assist the dynamical multiferroicity, enabling a combined ferroelectric and ferromagnetic switching at ultrafast time scales. It is worth stressing that the induced magnetic moment can be seen a quasi-static magnetization created on a timescale of a picosecond. This is about one order of magnitude faster than the fastest spin switching reported to date by S. Schlauderer *et al*.[38]. The applicability of the optical generation and control of magnetization by means of circular phonons can also be extended to the class of two-dimensional material in which chiral phonons exist intrinsically, such as transition metal dichalcogenides. These materials have recently been predicted and observed to host chiral phonons that are intrinsically circularly polarised[40,41]. Finally, we anticipate that our results will stimulate further research on the microscopic understanding of inertia and angular momentum transfer at ultrafast and atomic scales, and help further exploring and understanding intriguing entangled orders in condensed matter[42,43,44,45].

**Acknowledgements** M.B., A.V.B. and S.B. acknowledge support from the Knut and Alice Wallenberg Foundation, grants no. 2017.0158 and 2019.0068. A.V.B acknowledges the European Research Council ERC HERO-810451 synergy grant. T.T acknowledges support from JSPS KAKENHI grant number 21K03424. We gratefully acknowledge discussion with M. Geilhufe, M. Fechner, D. Afanasiev, V. Unikandanunni, and L. Benfatto.

**Author contributions** S.B. proposed and designed the experiment and the model, with contributions from M.B. and A.V.B.. M.B., M.P and S.B analysed the data presented in this work. M.B. and M.P. performed the experiments and wrote the code to simulate the coupled driven dynamics. M.U. performed simulations to calculate the nonlinear optical response. B.W. and T.T. designed and performed the *ab initio* calculations. M.C.H. performed additional independent measurements to confirm some of the experimental findings. All authors discussed the results. M.B. and S.B. wrote the manuscript with input from all authors.

## Methods

**Sample details.** The sample considered was a 10 mm x 10 mm, 500-$\mu$m-thick SrTiO$_3$ crystal substrate from MTI Corporation, with the [001] crystallographic direction normal to the cut direction. Both sides are polished.

**Experimental methods.** Broadband single-cycle terahertz radiation was generated by optical rectification in a DSTMS (dimethylamino-methyl-stilbazolium-trimethylbenzenesulfonate) crystal[15] of a 40 fs-long, 800 $\mu$J near infrared laser pulse centred at a wavelength of 1300 nm. This pulse was generated via optical parametric amplification from a 40 fs-long, 6.3 mJ pulse centred at 800 nm wavelength, produced by a 1 kHz regenerative amplifier. The terahertz pulses were focused into the sample by means of three parabolic mirrors to a rounded shape beam of approximately 0.5 mm in diameter. The exact size of the beam is not crucial to estimate the fluence since we characterised the electric field of the radiation. A pair of wire-grid polarisers was used to tune the field amplitude without affecting the pulse shape. The narrow band terahertz radiation was obtained by filtering the broadband field by means of a 3 THz band-pass filter resulting in a peak frequency of 3 THz and a full width at half maximum of 0.5 THz.

The probe beam was a 40 fs-long pulse at 800 nm wavelength, produced by the same 1 kHz regenerative amplifier used to generate the pump radiation. A beta-BBO crystal was used to convert the probe wavelength to 400 nm. The probe polarisation was set by means of a nanoparticle linear film polariser. The probe size at the sample was approximately 100 $\mu$m in diameter, substantially smaller than the terahertz pump.

To record the change in the polarisation state of the probe beam, a Wollaston prism was used to implement a balanced detection scheme with two photodiodes. Before the Wollaston, a half-wave plate was used for detecting the Kerr rotation . The signals from the photodiodes were fed to a lock-in amplifier, whose reference frequency (500 Hz) came from a mechanical chopper mounted along the pump path.

**Characterisation of the terahertz electric field.** The electric field component of the terahertz pulse at the sample location was characterised by means of electro-optical sampling[46] in a 50-$\mu$m-thick (110) cut GaP crystals. The maximum measured terahertz peak electric field was approximately 1.15 MV cm$^{-1}$. The peak frequency of the terahertz pump pulse is at 2.7 THz with measurable components extending up to approximately 5 THz. After the filtering the field with the 3THz band-pass filter, a typical measured terahertz peak electric field was around 200-300 kV cm$^{-1}$. The sampled terahertz pump traces are reported in Extended Data Figure 1, with the narrowband data representing the field used for measuring all of the data in the main text.

**Circular polarisation of the terahertz beam**. After filtering the broadband terahertz radiation with a terahertz band-pass filter, the linear polarisation state can be converted into circular by using a terahertz quarter-wave plate. For this purpose, we chose a Tydex quarter-wave plate made of x-cut terahertz-grade crystal quartz, whose thickness is adjusted to provide a $\pi/2$ phase shift at 3 THz. However, for waveplates the phase shift is very sensitive to the radiation frequency, and we are not working with a monochromatic beam. Nonetheless, in our case the use of a quarter-wave plate is justified by the fact the bandwidth of the filtered pulse is narrow enough to allow the waveplate to operate according to its design.

**Evaluation of complex refractive index.**
The complex reractive index $\tilde{n} = n + ik$ of SrTO$_3$ was derived from a combination of previous ellipsometry measurements on STO thin films[46] and hyper-Raman scattering in bulk STO[18]. In particular,

$$n = \sqrt{\frac{|\tilde{\varepsilon}| + \varepsilon_1}{2}}, \; k = \sqrt{\frac{|\tilde{\varepsilon}| - \varepsilon_1}{2}}$$

where $\tilde{\varepsilon} = \varepsilon_1 + i\varepsilon_2$ is the complex permittivity. In order to estimate the permittivity values in the experimental temperature range 160 K < T < 375 K for our specific sample, we first used the experimental data of Ref.[47] at $T$ = 300 K (see Extended Data Figure 2) which contained the broadband response that can be fitted with a Lorentz oscillator. Then, in order to adjust it to our case, we rigidly shifted the curve, moving the peak from 3 THz to 2.7 THz, in accordance to our own data and Ref.[17] on bulk samples. All center frequency and linewidth values at the different temperatures are listed in Extended Data Table 1. All values for n and k are reported in Extended Data Table 2.

**Modelling the terahertz reflectance, transmittance and absorptance.** The electric field reflection, absorption and transmission properties have been calculated in an air / STO / air stack by means of the analytical formulas for optical trilayers at normal incidence[48]:

$$r = \frac{C - C \exp(2i\delta)}{1 - C^2 \exp(2i\delta)}, \qquad t = \frac{\frac{4n}{(1+n)^2} \exp(i\delta)}{1 - C^2 \exp(2i\delta)},$$

$$C = \frac{1-n}{1+n}, \qquad \delta = \frac{2\pi d}{\lambda} n,$$

Where $n$ is the refractive index of STO, $d$ is the thickness of the STO sample, $\lambda$ is the wavelength, and the refractive index of air is considered to be 1. The reflectance, transmittance and absorptance are given, respectively, by $R = |r|^2$, $T = |t|^2$ and $A = 1 - R - T$. Considering $d = 500$ μm, $\lambda = 100$ μm (3 THz) and $n = 3.8 + i6.4$, from Ref.[17] at a temperature of 300 K, we get $R \approx 0.76, T \approx 0, A \approx 0.24$. Since $T \approx 0$, it is also interesting to estimate the decay length $l_{decay}$ of the electric field inside STO, to know how much the pump radiation penetrates into the sample, which is

$$l_{decay}(n) = \frac{\lambda}{2\pi \Im(n)} \approx 2.49 \text{ μm}.$$

The estimated penetration depth in the experimental temperature range $160K < T < 375K$ is listed in Extended Data, Table 2.

**Polarization rotation and magnetic field estimates.** The measured probe polarization rotation allows for calculating the magnetic field induced in STO. According to theory, Faraday rotation $\vartheta_F$ and magnetic field are connected via the equation[20]

$$\vartheta_F = VB \int_0^d \exp\left(-2\frac{z}{l_{decay}}\right) dz = VB \frac{l_{decay}}{2},$$

since $l_{decay} \ll d$, where $d$ is the STO thickness. The parameter $B$ represents the amplitude of the magnetic field at the surface, $l_{decay}$ is the decay length of the pump field and $V$ is the Verdet constant. The factor of 2 in the exponential function appears due to the fact that the induced magnetic field is proportional to the square of the pump electric field. To extract the magnetic moment generating $\vartheta_F$, we exploited the relation $B = \mu_0 M$, where $M$ is then the magnetization induced by the pump. Considering the STO lattice parameter $a = 3.9$ Å, the magnetic moment per unit cell $\mu$ is given by

$$\mu = \frac{2\vartheta_F a^3}{\mu_0 V l_{decay}}.$$

Even if the measurements reported in the main text are performed in reflection (Kerr rotation $\vartheta_K$), we evaluate the magnetic field considering a transmission measurement (Faraday geometry), since we found in literature a reliable value for the Verdet constant only for that case. To this end, we measured the Faraday rotation during the same set of experiments described in the main text, and it turned out to be comparable with the Kerr rotation within a factor of 2. Moreover, the pump penetration depth still is the limiting factor for the decay length $l_{decay}$ to be considered in the above equation, since even in reflection the STO thickness probed by the probe pulse results to be bigger than the pump penetration depth. The thickness contributing to the probe signal in reflection for an ultrafast pulse can be estimated through the distance travelled in the material during the pulse duration. For a 400 nm probe pulse ($n_{STO} = 2.6$, from Ref.[49]) of 50 fs duration, the STO thickness travelled during that interval corresponds to approx. 5.8 μm, which is bigger than all the pump field penetration depths listed in Extended Data, Table 2. According to Ref.[50] we have $V \approx 250$ rad m$^{-1}$T$^{-1}$, so that for $\vartheta_K = 10$ μrad and $l_{decay} = 2.49$ μm, the magnetic field at the surface turns out to be $B \approx 0.032$ T. The average energy $\epsilon$ stored per unit surface in such a magnetic field is given by:

$$\epsilon = \frac{1}{2}\frac{1}{2\mu_0} B^2 \int_0^d \exp\left(-4\frac{z}{l_{decay}}\right) dz = \frac{1}{16\mu_0} B^2 l_{decay} \approx 0.013 \text{ μJ cm}^{-2}.$$

The first factor of $1/2$ in the definition of $\epsilon$ is due to the time average of the square of a sine wave, since we approximate the slow oscillation in Fig. 3(b) of the main text with a sinusoidal function. The integral takes into account the fact that the induced magnetic field does not fill the whole sample volume, but it has a finite penetration depth. The energy $\epsilon$ is delivered by the pump pulse, and its fluence can be calculated by integrating the square of the trace shown in Extended Data Figure 1, obtaining a result of approximately 60 µJ cm$^{-2}$, much higher than the energy per unit surface in the generated magnetic field.

**Ab-initio calculations.** First-principles phonon calculations of cubic SrTiO$_3$ were performed within density functional theory (DFT) by using the *Vienna Ab initio Simulation Package* (VASP)[51], which implements the projector augmented wave (PAW) method[52]. The adopted PAW potentials treat Sr 4s$^2$ 4p$^6$ 5s$^2$, Ti 3s$^2$ 3p$^6$ 4s$^2$, and O 2s$^2$ 2p$^2$ as the valence states. Cut-off energy of 550 eV was used, and the Brillouin-zone integration was performed with the 12×12×12 gamma-centred $k$-point mesh. For an accurate description of the phonon potential energy surface, the Heyd–Scuseria–Ernzerhof hybrid functional (HSE06[53]) was adopted. The lattice constant was optimized within the HSE06; the optimized value of 3.900 Å agrees well with the experimental value, 3.905 Å.

The calculation of the effective phonon frequencies and eigenvectors at room temperature were conducted based on the self-consistent phonon (SCP) theory, as implemented in the ALAMODE software[54]. The harmonic and fourth-order interatomic force constants (IFCs), which are necessary as inputs to the SCP calculation, were calculated by using the real-space supercell approach with the 2×2×2 supercell. The harmonic IFCs were estimated by systematically displacing each atom in the supercell from its equilibrium site by 0.01 Å, calculating forces by DFT, and fitting the harmonic potential to the displacement-force datasets. The fourth-order IFCs were estimated by using the compressive sensing method, where 40 training structures were generated by combining DFT-MD with random displacements following the prescription of Ref.[55].

After obtaining the harmonic and anharmonic IFCs, the effective phonon frequency $\omega$ with branch index $\nu$ at wavevector $q$ and the corresponding eigenvector are obtained by solving the self-consistent the SCP equation

$$\omega_{q\nu}^2(T) = \left(C_q^\dagger \Lambda_q^{HA} C_q\right)_{\nu\nu} + \frac{1}{2}\sum_{q',\nu'} \Phi^{SCP}(-q\nu; q\nu; -q'\nu'; q'\nu') \times \frac{\hbar\left(1 + 2n(\omega_{q'\nu'})\right)}{2\omega_{q'\nu'}},$$

where $\Lambda_q^{(HA)} = diag(\widetilde{\omega}_{q,1}^2, \ldots, \widetilde{\omega}_{q,\nu}^2)$ with harmonic frequencies $\widetilde{\omega}$. $C$ is a unitary transformation matrix that modifies the polarisation vector at finite temperature, $\Phi^{SCP}$ is the fourth-order anharmonic force constants, and $n_{q,\nu}(\omega_{q,\nu})$ is the Bose-Einstein distribution. The equation was solved numerically by iteratively updating the effective frequency $\omega_{q\nu}(T)$ and the unitary matrix $C_q$ for the phonon modes at the gamma-centred 2×2×2 $q$ points.

The summation over the $q'$ points was conducted with the denser 10×10×10 $q'$ points, which was sufficient to achieve convergence. The quartic coupling coefficient $\Phi^{SCP}(-q\nu; q\nu; -q'\nu'; q'\nu')$ was obtained from the fourth-order IFCs by using the Fourier interpolation. The LO-TO splitting was considered in the SCP calculation. The obtained SCP frequencies agree well with the inelastic neutron scattering data, as shown in the Extended Data Figure 3.

The anharmonic coupling coefficients of the triply-degenerate $\Gamma_{15}$ modes at room temperature were obtained by transforming the anharmonic IFCs into the normal coordinate basis. In this study, the anharmonic coupling terms up to the fourth-order were included in $V(Q_1, Q_2)$. The normal coordinate $Q_\nu$ at finite temperature is given as $Q_\nu = \sum_\kappa \sqrt{m_\kappa} e_\nu(\kappa) \cdot u(\kappa)$, where $m_\kappa$ is the mass of atom $\kappa$ and $u^\alpha(\kappa)$ is its displacement in the $\alpha$ direction. The polarisation vector at room temperature, $e_\nu(\kappa)$, was calculated as $e_\nu(\kappa) = \sum_{\nu'} \tilde{e}_{\nu'}(\kappa) [C_{q=0}]_{\nu'\nu}$, where $\tilde{e}_\nu(\kappa)$ is the harmonic polarisation vector and $C_q$ is the unitary matrix obtained as a solution to the SCP equation. Since the polarisation mixing is significant in SrTiO$_3$, the temperature-dependence of the polarisation vectors is noteworthy, as shown in the Extended Data Table 3 for the 300 K case. Since each atomic site of cubic SrTiO$_3$ is an inversion centre, all cubic coefficients became exactly zero. The effective charges of the $\Gamma_{15}$ modes were also calculated as

$$Z_{\nu,\alpha}^* = \sum_{\kappa\beta} Z_{\kappa,\alpha\beta}^* \frac{e_\nu^\beta(\kappa)}{\sqrt{m_\kappa}}$$

with $Z_{\kappa,\alpha\beta}^*$ being the Born effective charge of atom $\kappa$, and they are reported in Extended Data Table 4 for the 300 K case.

**Phenomenological model for anharmonically coupled oscillators.** In order to model the driven circular excitation of the ferroelectric mode along the two orthogonal directions, we derive the effective phonon potential for two of the three-fold degenerate modes at $q = 0$, where the anharmonic coupling is included up to fourth order, i.e.

$$V(Q_1, Q_2) = \frac{1}{2}\omega^2 Q_1^2 + \frac{1}{2}\omega^2 Q_2^2 + \frac{1}{4} k Q_1^4 + \frac{1}{4} k Q_2^4 + \chi Q_1^2 Q_2^2 + \psi Q_1^3 Q_2 + \psi Q_1 Q_2^3,$$

where $Q$ is the normal coordinate in the real space, the indexes 1 and 2 refer to the two degenerate branches of the soft phonon along [100] and [010], $k$ is the anharmonic contribution to the potential, and $\chi$ and $\psi$ are the phonon-phonon coupling terms. Since $Q_1$ and $Q_2$ are orthogonal to each other and the phonon potential spanned by them has a $C_4$ symmetry, $\psi = 0$. The resulting potential, with calculated parameters stated in Extended Data Table 1, is the one representing two coupled anharmonic oscillators. The solution of this model is obtained by numerical integration of its equation of motion

$$\ddot{Q}_i + \frac{\partial V}{\partial Q_i} + \Gamma \dot{Q}_i = Z^* \tilde{E}_i^{THz}, \quad i=1, 2,$$

where $\Gamma$ accounts for the lifetime of phonons and $Z_i \tilde{E}_i^{THz}$ is the oscillator coupling to the driving field via the mode effective charge $Z^*$. The effective field in the sample is expressed through the term $\tilde{E}_i^{THz} = \alpha E_i^{THz}$ where $\alpha$ quantify the amount of field actually experienced (not screened) by the sample. The value of $E_i^{THz}$ was fixed from our experiment, while the values of $\Gamma$ and $\omega$ at room temperature were taken from hyper-Raman measurements on bulk STO[17]. Finally, the induced magnetic moment can be calculated using the polarisation $P = ZQ$, and reads

$$\mu = \gamma Q \times \dot{Q} = \sum_i \gamma_i Q_i \times \dot{Q}_i = \sum_i \gamma_i L_i,$$

where $i$ now represents the i-th atom in the unit cell (Sr, Ti, O, O, O), $\gamma_i = e Z_i^*/2m_i$ is a gyromagnetic ratio and $L_i = Q_i \times \dot{Q}_i$ is an angular momentum. The calculated magnetic moment per unit cell $\mu$ is shown in Fig. 4a in the time domain, and in Fig. 4b in the frequency domain using the identical approach used to process the experimental data. All parameters used to solve the equation of motion were fixed, except for $\alpha$ which was adjusted to the value of 0.7. The mode effective charge $Z^*$ and potential parameters $k$, $\chi$ and $\psi$ have been calculated from first principles (see Extended Data Table 1), while the excitation field $\tilde{E}_i^{THz}$ and the phonon frequency $\omega$ and lifetime $\Gamma$ are those obtained in experiments[17].

**Extended Data**

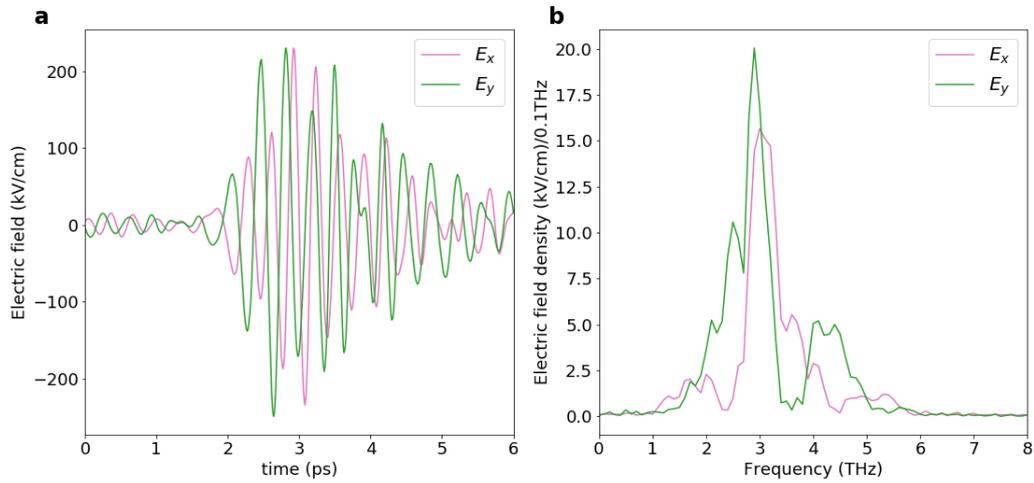

**Extended Data Figure 1 | Electro-optical sampling data of the circularly polarized terahertz field**. (a) Recorded temporal trace and (b) its fast Fourier transform. The measurement was performed using a 50 μm thick GaP crystal cut along the 110 crystallographic direction, and the broadband pulse was filtered by means of a 3THz filter with approximately 10% bandwidth, as described in the Methods. The two orthogonal components of the electric field are shown with pink ($E_x$) and green ($E_y$) solid lines. In panel (a) the reflected terahertz field from the back side of the GaP crystal starts to interfere with the direct beam at approximately $t = 3.8$ ps. This complicates the electro-optic sampling data presented here, but it has no importance in all measurements on the STO crystal presented in the main text. In STO, the large terahertz absorption suppresses the back side reflection below the experimental noise level.

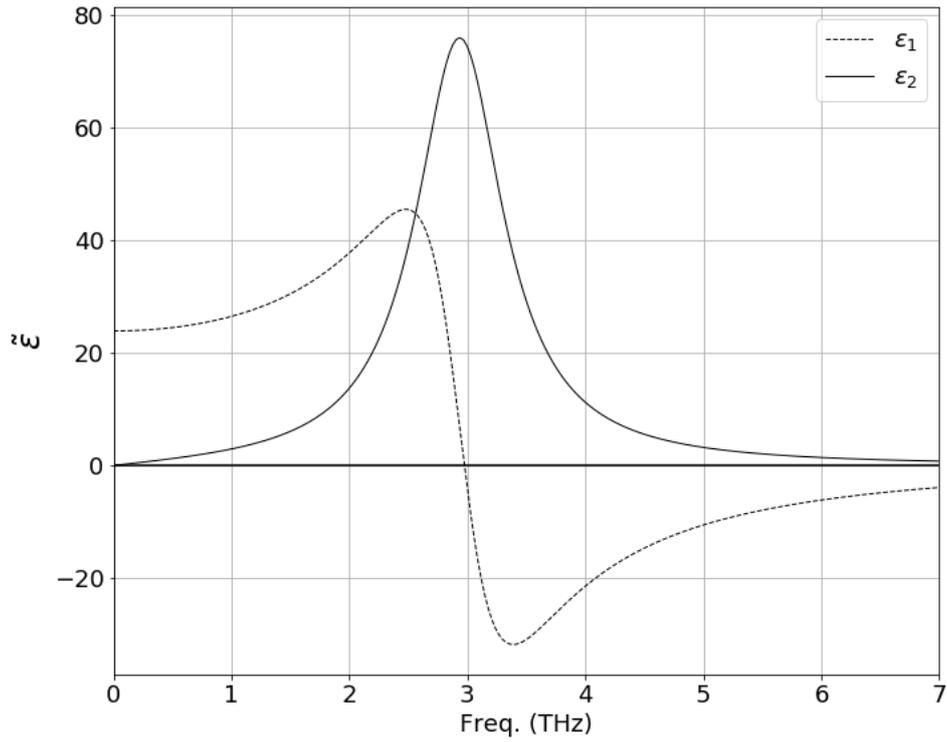

**Extended Data Figure 2 | Complex permittivity of STO at T=300K used for the simulations.** The functional dependence was found assuming a Lorentz oscillator with linewidth and oscillator strength found in Ref. 44, while the resonance frequency was matched with the experimental soft phonon frequency of Ref. 17 and listed in Extended Data Table 4.

| T (K) | $\omega_0/2\pi$ (THz) | $\gamma/2\pi$ (THz) | k (THz²Å⁻²AMU⁻¹) | χ (THz²Å⁻²AMU⁻¹) | $Z^*$ (e⁻AMU⁻¹ᐟ²) | g (1/e⁻) |
|---|---|---|---|---|---|---|
| 160 | 1.85 | 0.35 | 1556 | 127 | 1.62 | 0.057 |
| 180 | 2.00 | 0.41 | 1496 | 125 | 1.61 | 0.058 |
| 200 | 2.15 | 0.48 | 1436 | 123 | 1.60 | 0.059 |
| 220 | 2.27 | 0.52 | 1376 | 121 | 1.59 | 0.060 |
| 240 | 2.40 | 0.55 | 1312 | 119 | 1.58 | 0.061 |
| 260 | 2,48 | 0.58 | 1428 | 117 | 1.57 | 0.062 |
| 280 | 2.60 | 0.61 | 1180 | 115 | 1.56 | 0.062 |
| 300 | **2.70** | 0.63 | 1112 | 113 | 1.54 | 0.063 |
| 320 | 2.80 | 0.69 | 1044 | 110 | 1.52 | 0.064 |
| 340 | 2.90 | 0.74 | 976 | 108 | 1.50 | 0.064 |
| 360 | 3.05 | 0.77 | 908 | 105 | 1.49 | 0.065 |
| 375 | 3.10 | 0.81 | 840 | 102 | 1.47 | 0.065 |

**Extended Data Table 1 | Phonon parameters used to solve the coupled oscillator equation.** The phonon center frequency $\omega_0$ and linewidth $\gamma$ are taken from the experimental value in Ref. 17. The nonlinearity coefficient k, the nonlinear coupling coefficient χ, the effective charge $Z^*$ and the gyromagnetic ratio g are taken from *ab initio* calculations.

| T (K) | n | k | $l_{decay}$ (µm) |
|---|---|---|---|
| 160 | 1.03 | 4.42 | 3.60 |
| 180 | 1.2 | 4.72 | 3.37 |
| 200 | 1.49 | 5.07 | 3.14 |
| 220 | 1.77 | 5.38 | 2.96 |
| 240 | 2.18 | 5.73 | 2.78 |
| 260 | 2.5 | 5.94 | 2.68 |
| 280 | 3.12 | 6.23 | 2.56 |
| 300 | 3.78 | 6.40 | 2.49 |
| 320 | 4.55 | 6.42 | 2.48 |
| 340 | 5.40 | 6.20 | 2.57 |
| 360 | 6.51 | 5.36 | 2.97 |
| 375 | 6.76 | 4.98 | 3.19 |

**Extended Data Table 2 | Real and imaginary parts of the refractive index in STO, and corresponding penetration depth $l_{decay}$.** All values are estimated considering the dielectric function plotted in Extended Data Figure 1 and properly shifted to take into account the variation of the dielectric function with temperature.

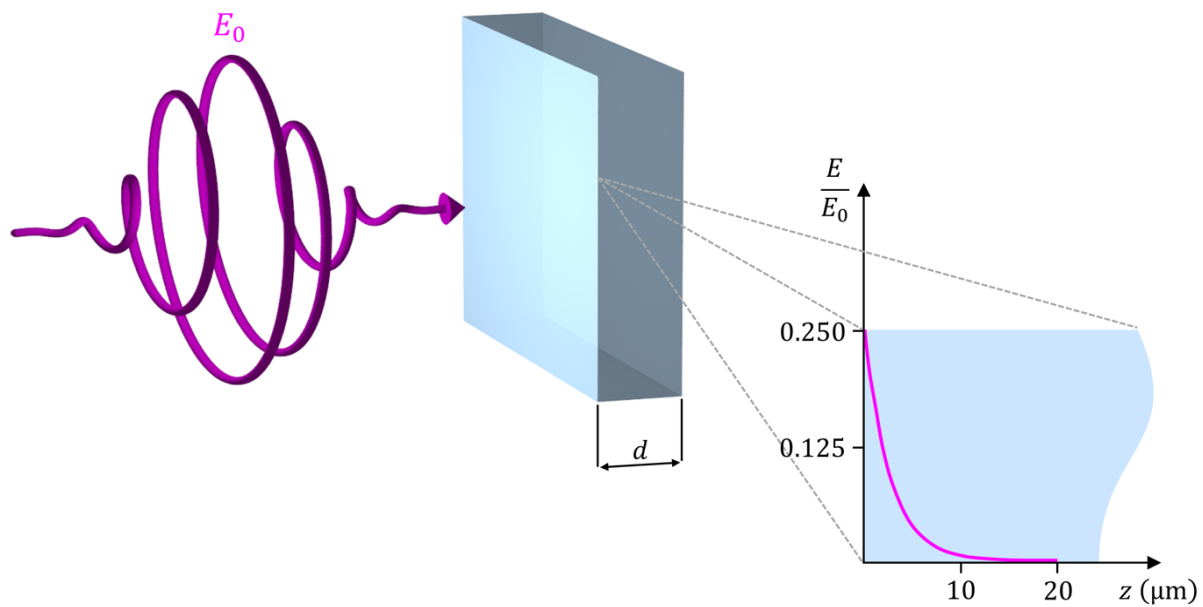

**Extended Data Figure 2 | Propagation of the terahertz field in the STO sample.** Schematic representation of the interaction between the circularly polarized terahertz pump pulse $E_0$ and the STO sample of thickness $d$. The plot shows the exponential decay of the electric field $E$ in the first few μm below the STO surface for $\tilde{n} = 3.78 + i6.40$ and $\omega = 3$ THz.

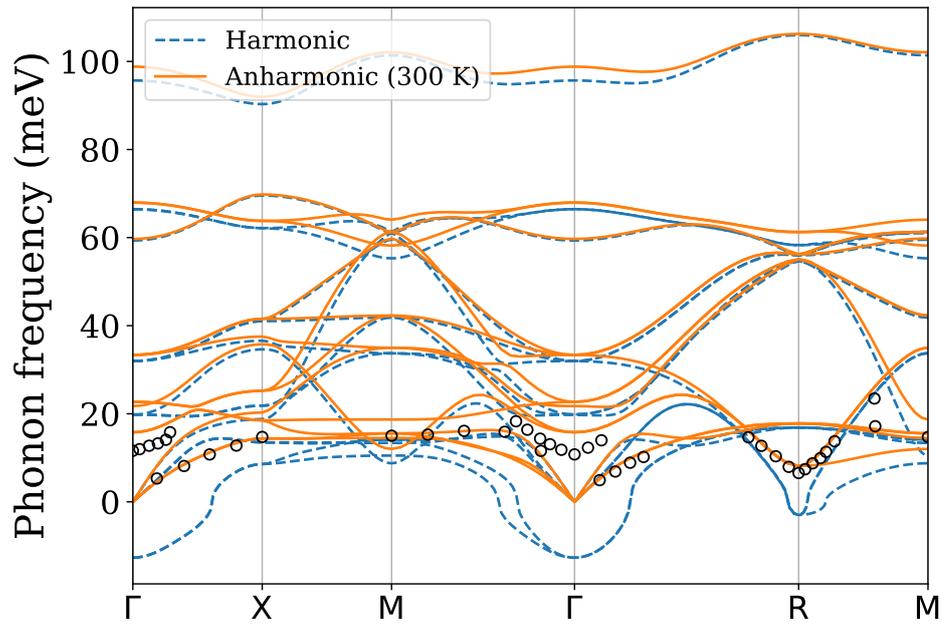

**Extended Data Figure 3 | Phonon dispersion curves of cubic SrTiO$_3$.** The curves were calculated based on the harmonic approximation (dashed lines) and the self-consistent phonon theory at 300 K (solid lines). The inelastic neutron scattering data[18] are also shown for comparison.

|  |  | Harmonic | | | Anharmonic ($T = 300$ K) | | |
| --- | --- | --- | --- | --- | --- | --- | --- |
|  | $\kappa$ | $\tilde{e}^x(\kappa)$ | $\tilde{e}^y(\kappa)$ | $\tilde{e}^z(\kappa)$ | $e^x(\kappa)$ | $e^y(\kappa)$ | $e^z(\kappa)$ |
| Mode $Q_1$ | | | | | | | |
|  | Sr | -0.191 | 0 | 0 | -0.415 | 0 | 0 |
|  | Ti | -0.552 | 0 | 0 | -0.299 | 0 | 0 |
|  | O1 | 0.511 | 0 | 0 | 0.483 | 0 | 0 |
|  | O2 | 0.445 | 0 | 0 | 0.503 | 0 | 0 |
|  | O3 | 0.445 | 0 | 0 | 0.503 | 0 | 0 |
| Mode $Q_2$ | | | | | | | |
|  | Sr | 0 | -0.191 | 0 | 0 | -0.415 | 0 |
|  | Ti | 0 | -0.552 | 0 | 0 | -0.299 | 0 |
|  | O1 | 0 | 0.445 | 0 | 0 | 0.503 | 0 |
|  | O2 | 0 | 0.511 | 0 | 0 | 0.483 | 0 |
|  | O3 | 0 | 0.445 | 0 | 0 | 0.503 | 0 |
| Mode $Q_3$ | | | | | | | |
|  | Sr | 0 | 0 | -0.191 | 0 | 0 | -0.415 |
|  | Ti | 0 | 0 | -0.552 | 0 | 0 | -0.299 |
|  | O1 | 0 | 0 | 0.445 | 0 | 0 | 0.503 |
|  | O2 | 0 | 0 | 0.445 | 0 | 0 | 0.503 |
|  | O3 | 0 | 0 | 0.511 | 0 | 0 | 0.483 |

**Extended Data Table 3 | Polarisation vectors.** The polarisation vectors of the triply degenerate ferroelectric soft modes at $q = 0$ calculated based on the harmonic approximation and the self-consistent phonon theory at 300 K. The modes $Q_1$, $Q_2$, and $Q_3$ correspond to the displacements in the [100], [010], and [001] directions, respectively. The fractional coordinates of each atom are as follows: Sr (0, 0, 0), Ti (1/2, 1/2, 1/2), O1 (0, 1/2, 1/2), O2 (1/2, 0, 1/2), O3 (1/2, 1/2, 0).

| $\epsilon^\infty$ | Atomic effective charge $Z^*_{\kappa,\alpha\beta}$ (e) | | | | Mode effective charge $Z^*_{\nu,\alpha}$ ($e \cdot u^{-1/2}$) | |
| --- | --- | --- | --- | --- | --- | --- |
| | $Z^*$(Sr) | $Z^*$(Ti) | $Z^*$(O)$_\perp$ | $Z^*$(O)$_\parallel$ | $Z^*_{1,x}, Z^*_{2,y}, Z^*_{3,z}$ | Other terms |
| 4.984 | 2.553 | 6.704 | -1.941 | -5.375 | 1.5 | 0 |

**Extended Data Table 4 | List of the dielectric constant $\epsilon^\infty$ and the Born effective charges.** The values were calculated by using the HSE06 functional. $Z^*$(O)$_\perp$ ($Z^*$(O)$_\parallel$) represents the effective charge of an oxygen atom in the direction perpendicular (parallel) to the nearest titanium atom. The mode effective charges are calculated by using the polarisation vectors at 300 K obtained by the self-consistent phonon theory.